\documentclass[11pt,twoside,a4paper,cmspaper,final,collab]{cms-tdr}
\def\svnVersion{78d9832-D}\def\svnDate{2026/02/19}\def\cmsCernNoTag{CERN-EP-2025-244}\def\cmsCernDate{\today}\def\cmsMessage{Submitted to the Journal of High Energy Physics}
\begin{document}\cmsNoteHeader{SUS-23-006}

\newcommand{\epem}{\ensuremath{\Pep\Pem}\xspace}
\newcommand{\mpmm}{\ensuremath{\PGmp \PGmm}\xspace}
\newcommand{\epmmmp}{\ensuremath{\Pepm\PGmmp}\xspace}

\cmsNoteHeader{SUS-23-006}

\title{Search for \texorpdfstring{\PZpr}{Z'} bosons decaying into charginos in final states with two oppositely charged leptons and missing transverse momentum in \texorpdfstring{$\Pp\Pp$}{pp} collisions at \texorpdfstring{$\sqrt{s} = 13$\TeV}{13 TeV}}

\author[cern]{The CMS Collaboration}
\date{\today}

\abstract{Massive leptophobic \PZpr bosons decaying to a pair of charginos are searched for in proton-proton collisions at $\sqrt{s} = 13$\TeV, using data samples collected by the CMS experiment in 2016, 2017, and 2018, corresponding to a total integrated luminosity of 138\fbinv. The \PZpr bosons originate from an additional $U(1)^\prime$ gauge symmetry extended to the minimal supersymmetric standard model. The final state consists of two oppositely charged leptons and missing transverse momentum. The signal extraction is performed with a parametrized neural network. The measurements are found to be consistent with the standard model expectations. Upper limits are set on the \PZpr boson production cross sections as a function of the \PZpr and chargino masses. The analysis excludes \PZpr boson masses up to about 3.5\TeV for the specific case of \PZpr bosons decaying exclusively to charginos, with the charginos decaying to \PW bosons and neutralinos.}

\hypersetup{
pdfauthor={CMS Collaboration},
pdftitle={Search for Z' bosons decaying into charginos in final states with two oppositely charged leptons and missing transverse momentum in pp collisions at sqrt(s) = 13 TeV},
pdfsubject={CMS},
pdfkeywords={CMS, supersymmetry, Z' bosons}}

\maketitle 

\section{Introduction}
\label{sec:introduction}

Searches for a heavy neutral gauge boson \PZpr have been performed by the ATLAS and CMS Collaborations 
since the beginning of data taking at the CERN LHC. 
These searches  primarily focused on direct resonance decays into standard model (SM) particles: 
lepton pairs, jet pairs, and dibosons~\cite{ATLAS:2019erb, CMS:2021ctt, CMS:2019gwf, ATLAS:2020fry,CMS:2022pjv}. 
The most stringent limits so far have been set by the dilepton searches, the other channels imposing relatively mild constraints.

There are well-motivated theoretical frameworks beyond the SM (BSM), also consistent with the current experimental constraints, 
suggesting scenarios where resonances decay into new particles.  
One example is the $U(1)^\prime$-extended minimal supersymmetric SM (UMSSM) scenario~\cite{Corcella:2012dw,Hewett:1988xc,Langacker:2008yv,Araz:2017wbp}, derived from embedding the minimal supersymmetric SM (MSSM)~\cite{Martin:1997ns} in the unified gauge group $E_6$. 
This scenario offers several advantages: 
a scalar singlet introduced in the model can induce both supersymmetry (SUSY)~\cite{Martin:1997ns} breaking and the dynamical generation of the $\mu$ parameter~\cite{Araz:2017wbp},
which would otherwise be introduced unnaturally at the electroweak scale in the MSSM. 
It also forbids rapid proton decay, without the need to introduce an ad hoc discrete $R$-parity symmetry. 

The $U(1)^\prime$ extension leads to an extra gauge boson, the \PZpr. 
In certain charge scenarios, this leads to leptophobic \PZpr bosons, 
suppressing the decays of the \PZpr boson to leptons~\cite{Araz:2017wbp, Frank:2020kvp, Frank:2020pui} 
and circumventing the stringent experimental constraints set by dilepton resonance searches. 
Instead, the \PZpr boson decays are dominated by other channels, 
including decays to SM final states with 
quark pairs, 
boson pairs (such as $\PWp\PWm$ or $\PH\PZ$), 
neutrino pairs, 
or SUSY final states with chargino or neutralino pairs, where charginos (neutralinos) are the mass eigenstates of the charged (neutral) superpartners of the SM gauge and Higgs bosons.

\begin{figure}[ht]
\centering
\includegraphics[width=0.4\textwidth]{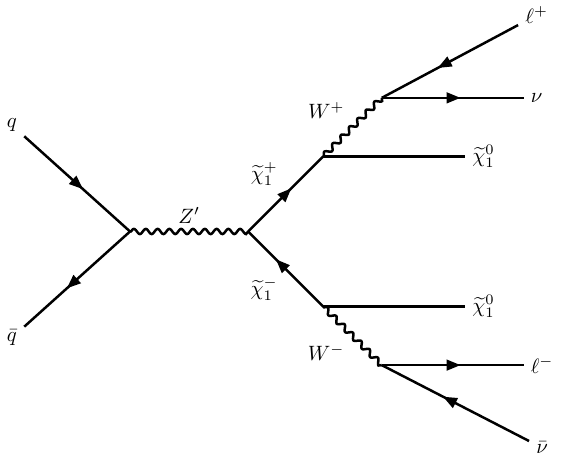}
\caption{Diagram showing the signal process studied in this analysis: 
a leptophobic \PZpr boson decaying into two charginos, each subsequently decaying into a lepton, a neutrino, and a neutralino.}
\label{fig:sig_process}
\end{figure}

This paper reports a search for leptophobic \PZpr bosons that decay into two charginos, $\PSGcpmDo$, 
each of them decaying into a \PW boson and the lightest neutralino, $\PSGczDo$
(the lightest supersymmetric particle, which is a dark matter candidate and is stable under R-parity conservation, given the $U(1)^\prime$ gauge structure of the UMSSM scenario),
with the \PW bosons decaying leptonically. 
This process, illustrated in Fig.~\ref{fig:sig_process}, 
leads to a distinct signature with two oppositely charged high-momentum leptons (\epem, \mpmm, or \epmmmp) 
and high missing transverse momentum (\ptmiss). 
Previously reported searches have not addressed this specific scenario. 
For example, the \PZpr boson searches in the dilepton channel described in Refs.~\cite{ATLAS:2019erb, CMS:2021ctt} 
have neither applied an explicit requirement on \ptmiss nor considered the kinematic properties imposed by the intermediate charginos. 
Moreover, direct chargino pair production searches, such as those reported in Refs.~\cite{ATLAS:2022hbt, CMS:2024gyw}, 
have not explored the kinematic properties imposed by the assumption that they are produced in the decay of a heavy particle.

The leptophobic \PZpr boson production cross section 
is similar to that of the \PZpr bosons predicted in other models, such as the sequential SM or those from non-supersymmetric $E_6$ gauge groups~\cite{ATLAS:2019erb, CMS:2021ctt, CMS:2019gwf, ATLAS:2020fry}. 
In the full UMSSM scenario, 
including the complete SUSY particle content with their predicted properties, 
the branching fractions of the \PZpr boson decay into charginos 
and of the subsequent decays of the charginos into leptonically decaying \PW bosons
are both of the order of a few percent~\cite{Araz:2017wbp, Frank:2020kvp}. 
In order to extract such small signal from the dominant SM backgrounds, 
the analysis employs a multivariate analysis approach.

\section{The CMS apparatus and event samples}

The CMS apparatus is a multipurpose detector~\cite{CMS:2008xjf}
designed to trigger on and identify electrons, muons, photons,
and (charged and neutral) hadrons~\cite{CMS:2020uim,CMS:2018rym,CMS:2014pgm}.
A superconducting solenoid of 6\unit{m} internal diameter provides a magnetic field of 3.8\unit{T}.
Within the solenoid volume are the silicon pixel and strip tracker,
a crystal electromagnetic calorimeter (ECAL), and a brass and scintillator hadron calorimeter (HCAL),
each composed of a barrel and two endcap sections. 
Forward calorimeters extend the pseudorapidity coverage provided by the barrel and endcap detectors. 
Muons are measured in gas-ionization detectors 
(drift tubes, cathode strip chambers, and resistive plate chambers)
embedded in the steel flux-return yoke outside the solenoid.
Events of interest are selected using a two-tiered trigger system.
The first level (L1), composed of custom hardware processors,
uses information from the calorimeters and muon detectors to select events
at a rate of 100\unit{kHz} within a fixed latency of 4\mus~\cite{CMS:2020cmk}.
The second level,
consisting of a farm of processors running a faster version of the full event reconstruction software,
reduces the rate to around 1\unit{kHz}, before data storage~\cite{CMS:2016ngn,CMS:2024aqx}.
A more detailed description of the CMS detector, 
together with a definition of the coordinate system used and the relevant kinematic variables, 
can be found in Refs.~\cite{CMS:2008xjf,CMS:2023gfb}. 

The proton-proton ($\Pp\Pp$) event samples used in the analysis were collected 
at a center-of-mass energy of 13\TeV in 2016, 2017, and 2018, 
with integrated luminosities of 36.3, 41.5, and 59.8\fbinv, 
respectively~\cite{CMS-LUM-17-003,CMS-PAS-LUM-17-004,CMS-PAS-LUM-18-002},
leading to a total integrated luminosity of 138\fbinv.

The analysis also relies on Monte Carlo (MC) simulated event samples, 
to evaluate the SM backgrounds and extract the signal yields.
The signal events are generated for the exclusive process 
$\Pp\Pp \to \PZpr \to \PSGcpDo \, \PSGcmDo \to \PWp \PSGczDo \, \PWm \PSGczDo \to \Pellp \PGn \PSGczDo  \, \Pellm \PGn \PSGczDo$. 
The hard process is generated at leading order (LO) precision using the \MGvATNLO~v2.6.5 generator~\cite{Alwall:2014hca},
with the NNPDF3.1 LO parton distribution functions (PDFs)~\cite{Ball:2017nwa} 
and assuming the strong coupling constant $\alpS(m_\PZ) = 0.13$, where $m_\PZ$ is the mass of the Z boson. 
The masses of the \PZpr boson and of the chargino are scanned, respectively, 
from 1.7 to 4.1\TeV in steps of 400\GeV and from 345 to 1845\GeV in steps of 250\GeV.  
The neutralino mass is taken to be half of the chargino mass, which arises from the assumption of gaugino mass unification at the grand unified theory (GUT) scale~\cite{Martin:1997ns,ParticleDataGroup:2024cfk}. 
The events are generated in a theoretical framework that assumes a negligible \PZpr boson width. 
The effective cross sections are calculated by setting the 
$\PZpr \to \PSGcpDo \, \PSGcmDo \to \PWp \PSGczDo \, \PWm \PSGczDo$ branching fraction to 100\%.

The \POWHEG~v2~\cite{Nason:2004rx,Frixione:2007vw,Alioli:2010xd,Frixione:2007nw} event generator is used to simulate, 
at next-to-leading order (NLO) precision in perturbative quantum chromodynamics (QCD), the following SM background processes:
$\PQt \PAQt$ pair production;
single top quark production in association with a \PW boson, and \PW plus \PZ boson pair production.
The \MGvATNLO event generator is used to simulate,
at NLO precision,
these other SM background processes:
Drell--Yan (DY),
$\PW \PZ$,
$\PQt \PAQt \PW$,
$\PQt \PAQt \PZ$,
$\PW \PW \PW$, and
$\PW \PW \PZ$. 
For the SM backgrounds, the NNPDF3.1 NNLO PDF sets are used~\cite{Ball:2017nwa}.

To model the fragmentation and parton showering processes, 
the generated events are processed through \PYTHIA~\cite{Sjostrand:2014zea},
using version 8.240.
For multi-jet NLO samples, partons in the matrix element calculation are merged with the parton shower using the FxFx prescription~\cite{Frederix:2012ps}.
The MC samples are simulated with the CP2 (for the signal, consistent with the LO matrix element) and CP5 (for the SM backgrounds) underlying-event tunes~\cite{CMS:2019csb}.
The simulated events are processed through a detailed simulation of the CMS detector,
based on the \GEANTfour package~\cite{Agostinelli:2002hh}, 
using the same trigger and reconstruction algorithms as used to collect and process the data.

\section{Object reconstruction, identification, and selection}

The objects used in this analysis are reconstructed starting with a particle-flow (PF) algorithm~\cite{CMS:2017yfk},
which aims to reconstruct and identify each individual particle in an event, 
through an optimized combination of information from the various elements of the CMS detector. 
The energy of the photons is directly obtained from the ECAL measurements. 
The electron's energy is determined by combining 
the electron momentum at the primary interaction vertex (as determined by the silicon tracker), 
the energy of the corresponding ECAL cluster, 
and the energy sum of all bremsstrahlung photons spatially compatible with originating from the electron track. 
The energy of muons is obtained from the curvature of the corresponding tracks. 
The energy of each charged hadron is computed by combining the momentum measured in the tracker with the 
matching ECAL and HCAL energy deposits, 
corrected using the response function of the calorimeters to hadronic showers. 
Finally, the energy of neutral hadrons is obtained from the corresponding corrected ECAL and HCAL energies. 

Electrons are reconstructed by associating a track in the silicon layers with a cluster of energy in the ECAL. 
Their momentum is evaluated by combining the energy measurement in the ECAL with the momentum measurement in the tracker. 
Electrons from $\PZ \to \Pe \Pe$ decays, with transverse momentum $\pt \approx 45\GeV$, 
have a momentum resolution between 1.6\% and 5\%;
it is generally better in the barrel than in the endcaps 
and also depends on the bremsstrahlung radiation emitted by the electrons~\cite{CMS:2020uim,CMS-DP-2020-021,CMS:2024ppo}. 
Electrons are identified by an algorithm based on a boosted decision tree (BDT),
selecting a working point with identification efficiency approximately 90\%. 
Isolation variables~\cite{CMS:2020uim} are included in the BDT training. 
The electrons are required to have $\pt > 25\GeV$ and pseudorapidity $\abs{\eta} < 2.5$, 
excluding the $1.44 < \abs{\eta} < 1.57$ transition region, situated between the barrel and endcap detectors.
They are also required to have relative isolation, $I_\text{PF}^\text{rel}$, smaller than 0.08,
computed as the ratio between 
the sum of the \pt of all PF particles within a cone of radius $\Delta R = 0.3$ around the electron trajectory
and the electron \pt, with corrections applied to mitigate the contribution from pileup
(i.e., additional $\Pp\Pp$ interactions within the same or nearby bunch crossings).

Muons are measured in the $\abs{\eta} < 2.4$ range, by associating a track in the silicon tracker with a track in the muon detectors, and are required to have $\pt > 25\GeV$.
The single-muon trigger efficiency exceeds 90\% over the full $\eta$ range 
and the efficiency to reconstruct and identify muons is larger than 96\%. 
Muons with \pt up to 100\GeV have a relative transverse momentum resolution of 1\% in the barrel and 3\% in the endcaps~\cite{CMS:2018rym}.
Muons are identified by applying a set of quality criteria to the candidates,
using a working point with an identification efficiency of 95\%.
Their relative isolation, computed using a cone of radius 0.4, is required to be $I_\text{PF}^\text{rel} < 0.15$.

Jets are reconstructed from the PF particles using the infrared- and collinear-safe anti-\kt algorithm~\cite{Cacciari:2008gp, Cacciari:2011ma} 
with a distance parameter $\Delta R = 0.4$. 
The jet momentum is determined as the vectorial sum of all particle momenta in the jet.
Simulation studies show that this value is, on average, within 5--10\% of the momentum of the jet at the particle level (that is, prior to its detection), 
over the entire \pt spectrum and detector coverage. 
Pileup interactions 
increase the number of tracks and calorimetric energy deposits, thereby increasing the apparent jet momentum.

To mitigate this effect, tracks identified as originating from pileup vertices are discarded 
and an offset is applied to correct for remaining contributions~\cite{CMS:2020ebo}. 
The jet energy corrections are derived, from simulation studies, 
so that the measured (average) jet energy is identical to the particle-level jet energy. 
In situ measurements of the momentum balance in 
dijet, 
$\text{photon} + \text{jet}$, 
$\PZ + \text{jet}$, 
and multijet events 
are used to evaluate residual differences between the jet energy scale in measured and simulated events, 
so that appropriate corrections can be applied~\cite{Khachatryan:2016kdb}. 
Additional selection criteria are applied to each jet, 
to remove jets potentially dominated by instrumental effects or reconstruction failures~\cite{CMS:2020ebo}. 
Jets overlapping with selected leptons, within a cone of $\Delta R = 0.4$, are rejected.
To select true jets and reject instrumental noise, 
tight selection criteria are imposed on variables related to the energy fractions and relative multiplicities 
of the various PF candidate types clustered in the jet.
The selected jets must have $\pt > 30\GeV$ and $\abs{\eta} < 2.4$.

Some of the jets passing the above selection criteria are identified by the DeepCSV algorithm~\cite{CMS:2017wtu}
as originating from the hadronization of bottom quarks; they are referred to as \PQb-tagged jets.
The chosen working point of the algorithm has an average \PQb tagging efficiency of 70\%,
for a gluon and light flavor jet misidentification rate of 1\%~\cite{CMS:2017wtu}.

The missing transverse momentum vector \ptvecmiss (of magnitude \ptmiss) is computed, for each event,
as the projection, in the plane perpendicular to the beam axis,
of the negative vector sum of the momenta of all reconstructed PF objects~\cite{CMS:2019ctu}. 
The contribution from particles from pileup interactions, which degrades the \ptmiss resolution, 
is mitigated using the pileup per particle identification algorithm~\cite{Bertolini:2014bba}. 
Events with anomalous \ptmiss, originating from reconstruction or detector failures, are filtered out~\cite{CMS:2019ctu}.

\section{Analysis method and signal selection}

The events are selected using single-electron triggers with electron \pt thresholds of 27 or 32\GeV,
and single-muon triggers with muon \pt thresholds of 24 or 27\GeV,
depending on the year of data taking. 
The efficiencies for the single-electron (single-muon) triggers are about 80\% (90\%) at the \pt efficiency plateau~\cite{CMS:2021yvr, CMS:2020uim}.
The events are required to have exactly two oppositely charged leptons (electrons or muons),
the leading (trailing) lepton having $\pt > 80$ (40)\GeV. 
The dilepton invariant mass, $m_{\Pell\Pell}$, is required to be larger than 100\GeV, to suppress backgrounds with on-shell \PZ bosons.  
To eliminate most of the DY$+$jets and \ttbar backgrounds, \ptmiss is required to be larger than 100\GeV.
To further suppress the \ttbar background, events with \PQb-tagged jets are rejected.  
The events passing these criteria are then distributed in three classes, defined by their lepton flavor combinations: 
\epem, \mpmm, and \epmmmp. The selection is summarized in Table~\ref{tab:selection}. 

\begin{table}[ht]
\centering
\caption{Signal event selection and search channels.}
\label{tab:selection}
\begin{tabular}{lccc}
Selection variables  & \multicolumn{3}{c}{Selection criteria} \\
\hline
Number of leptons & \multicolumn{3}{c}{$=$2} \\
Opposite charge lepton pairs & \multicolumn{3}{c}{$=1$ ($\Pep\Pem$, $\PGmp\PGmm$, $\Pepm\PGmmp$)} \\ 
Leading lepton $\pt$ & \multicolumn{3}{c}{$>$80\GeV} \\
Trailing lepton $\pt$ & \multicolumn{3}{c}{$>$40\GeV} \\
Dilepton invariant mass ($m_{\Pell\Pell}$) & \multicolumn{3}{c}{$>$100\GeV}  \\
Missing transverse momentum (\ptmiss) & \multicolumn{3}{c}{$>$100\GeV} \\
Number of b jets & \multicolumn{3}{c}{$=$0} \\
\end{tabular}
\end{table}
 
These event selection criteria reject most of the SM backgrounds, while retaining a high signal selection efficiency. 
However, the remaining irreducible backgrounds, 
primarily from the $\ttbar$, $\PQt \PW$, and $\PW\PW$ processes, 
exhibit features that resemble those of the signal events, 
so that no single variable exists that effectively discriminates the signal from the backgrounds.
Therefore, a multivariate analysis approach is adopted,
combining information from the kinematical and topological features that characterize the signal and background processes.

The analysis employs a fully connected parametrized neural network (PNN), to overcome the limitations of traditional training approaches. When training a neural network (NN) to extract a signal model with free parameters, 
such as the particle masses, several approaches can be considered. 
One of the most common approaches is to train a dedicated NN for a selected set of individual mass points in the signal model. 
While this method is expected to provide an optimal performance for each of those specific mass points, 
it does not guarantee a good performance for other mass points.
The PNN approach~\cite{Baldi:2016fzo} addresses this limitation by including the free model parameters as inputs in the NN training.
In our specific analysis, the additional training parameters are the masses of the \PZpr boson and of the chargino.

The PNN learns nontrivial correlations among the relevant variables and provides a model that maps the input variables to signal- or background-like scores.
These PNN scores (ranging between 0 and 1) are then used in a binned likelihood fit that provides the results of the analysis.
The PNN training is performed using half of the simulated events (signal and all SM backgrounds considered in this analysis, randomly sampled) that pass the selection criteria described above, except for the \PQb-tagged jet veto requirement. 
Since the PNN score is not expected to be strongly affected by the \PQb-tagged jet requirement, 
the training is performed on \PQb-tagged jet inclusive events, thereby increasing the number of events available for training. 
The remaining half of the events, which are not used for training, are used for the validation of the PNN performance. 
The training is done independently for each of the search channels, 
taking into account the cross section of each process along with event reweighting, such that the effective event yields of each signal and the total background are balanced.

The PNN training uses the variables listed below. 
The variables are selected because they exhibit a good discriminating power between the signal and the SM background processes.
For example, the signal features higher average lepton \pt values because of the high mass of the \PZpr boson 
and higher \ptmiss because the final state includes two massive neutralinos.
\begin{itemize}
\item The \pt of the leading ($\Pell_1$) and trailing ($\Pell_2$) leptons.
\item The dilepton invariant mass, $m_{\Pell\Pell}$.
\item The \ptmiss.
\item The magnitude of the vectorial sum of the dilepton \pt and $\ptvecmiss$, $\abs{ \ptvec(\Pell\Pell) + \ptvecmiss }$. 
\item $\abs{ \ptvec(\Pell\Pell)+\ptvecmiss } - \pt(\Pell\Pell)$, where $\pt(\Pell\Pell)$ is the magnitude of the dilepton \pt.
\item The transverse mass,
\begin{equation}
\mT = 2 \, \pt(\Pell\Pell) \, \ptmiss \, [1 - \cos\Delta\phi(\vec{\Pell\Pell}, \ptvecmiss)],
\end{equation}
where $\Delta\phi(\vec{\Pell\Pell}, \ptvecmiss)$ is the azimuthal distance between the dilepton and $\ptvecmiss$.
\item The ``stransverse mass"~\cite{Lester:1999tx,Barr:2003rg},
\begin{equation}
M_\mathrm{T2} = \min_{\ptvecmiss{}^{(1)} + \ptvecmiss{}^{(2)} = \ptvecmiss}
  \left[ \max \left( \mT(\Pell_1, \ptvecmiss), \mT(\Pell_2, \ptvecmiss) \right) \right],
\label{eq:MT2}
\end{equation}
where $\ptvecmiss{}^{(1,2)}$ are the vector \pt of two invisible objects whose momentum sum is \ptvecmiss and
$\mT(\Pell_{1,2}, \ptvecmiss)$ are the transverse masses obtained by pairing $\ptvecmiss{}^{(i)}$ with either of the two leptons.
The stransverse mass is a discriminating variable particularly useful in events with more than one invisible particle in the final state. For example, in \ttbar events that decay leptonically, $M_\mathrm{T2}$ exhibits an endpoint at the \PW boson mass. 
\end{itemize}

These variables are complemented with variables representing angular distances between the various event objects.
In the azimuthal dimension, we have the distances 
between the two leptons, $\Delta\phi(\Pell_1,\Pell_2)$, 
between $\ptvecmiss$ and the dilepton, $\Delta\phi(\ptvecmiss,\ptvec(\Pell_1\Pell_2))$,
and between $\ptvecmiss$ and each of the two leptons, $\Delta\phi(\ptvecmiss,\ptvec(\Pell_1))$ and $\Delta\phi(\ptvecmiss,\ptvec(\Pell_2))$.
For each of these variables, the signal distribution peaks at $\pi$, while the background processes have flatter distributions. 
There is also the angular distance in $\eta$ and azimuth, $\Delta R = \sqrt{\smash[b]{ (\Delta \phi)^2 + (\Delta \eta)^2 }}$, between the two leptons, $\Delta R(\Pell_1,\Pell_2)$.
The signal distribution peaks close to 3 and reaches higher values, while the background processes have broader distributions and mostly populate lower values.

\begin{figure}[ht]
\centering
\includegraphics[width=0.4\textwidth]{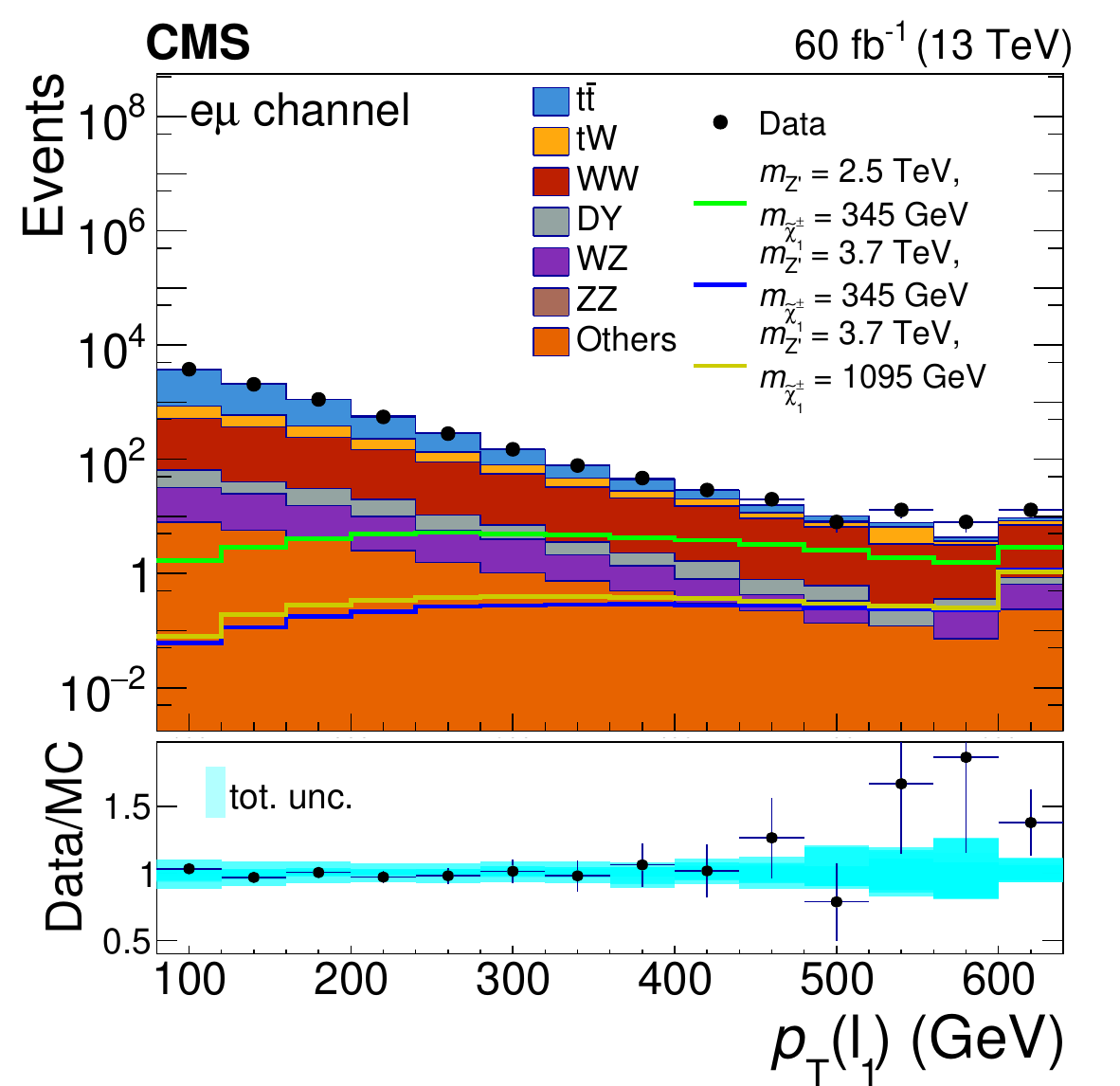}
\includegraphics[width=0.4\textwidth]{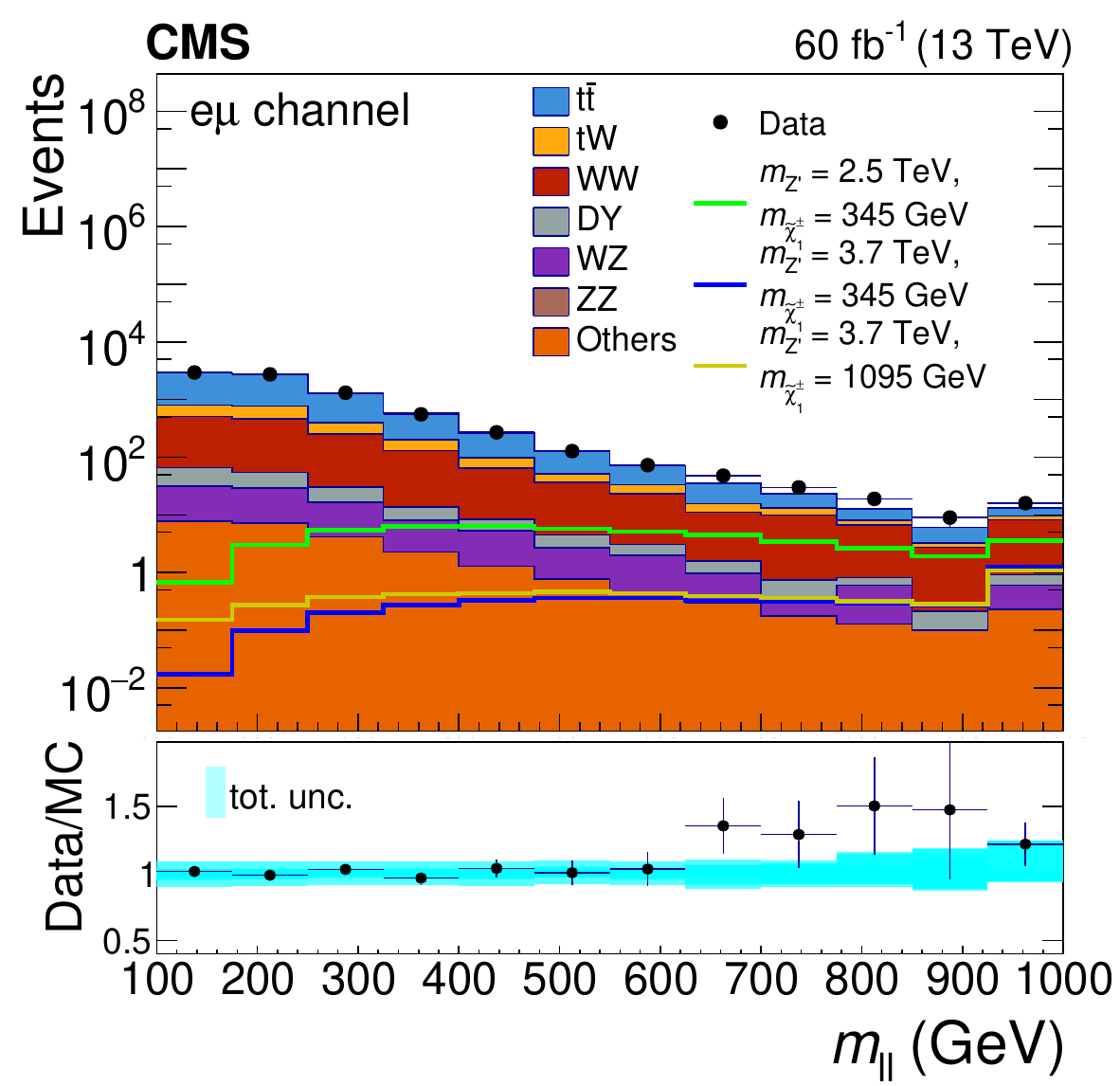}
\includegraphics[width=0.4\textwidth]{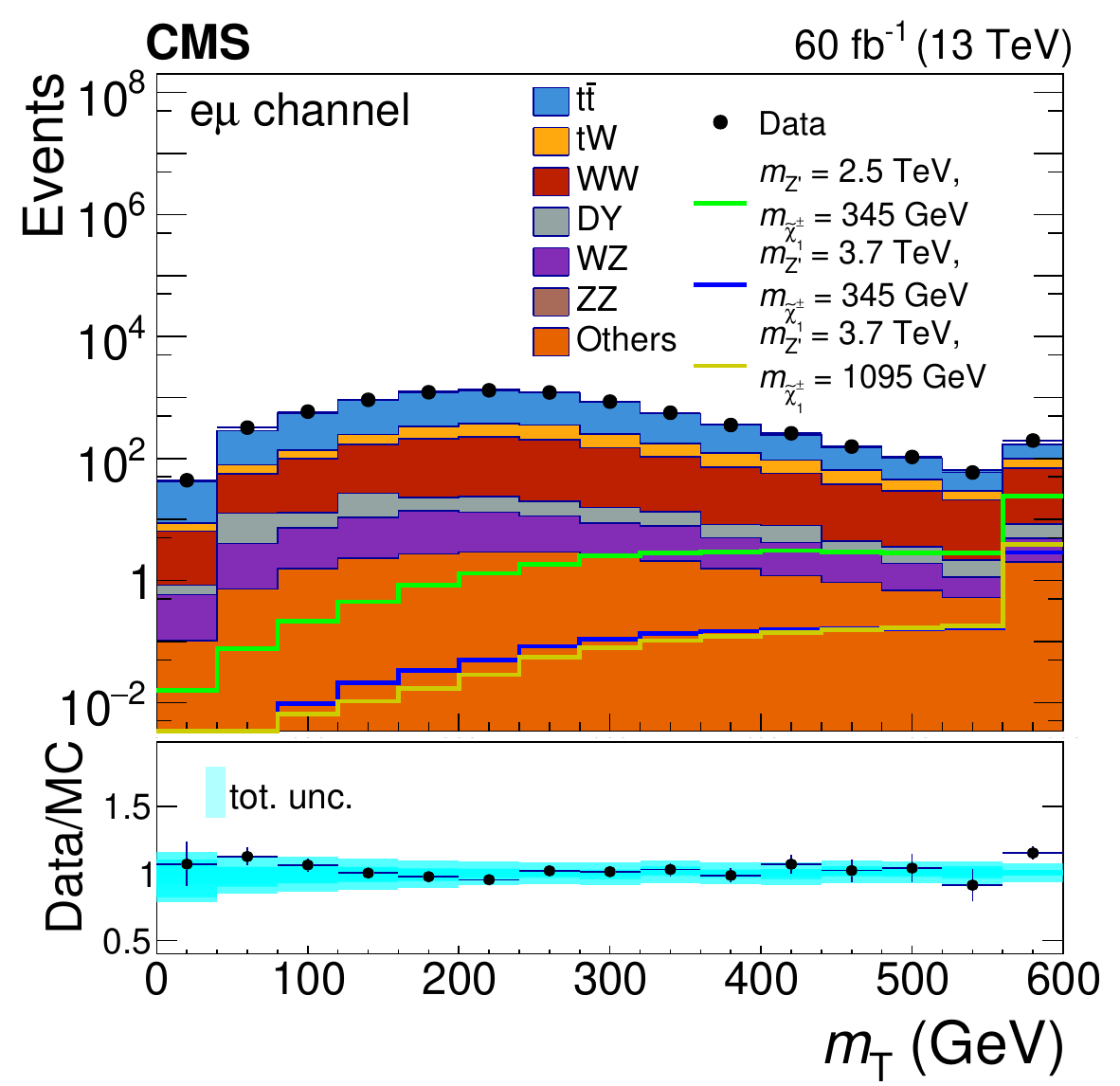}
\includegraphics[width=0.4\textwidth]{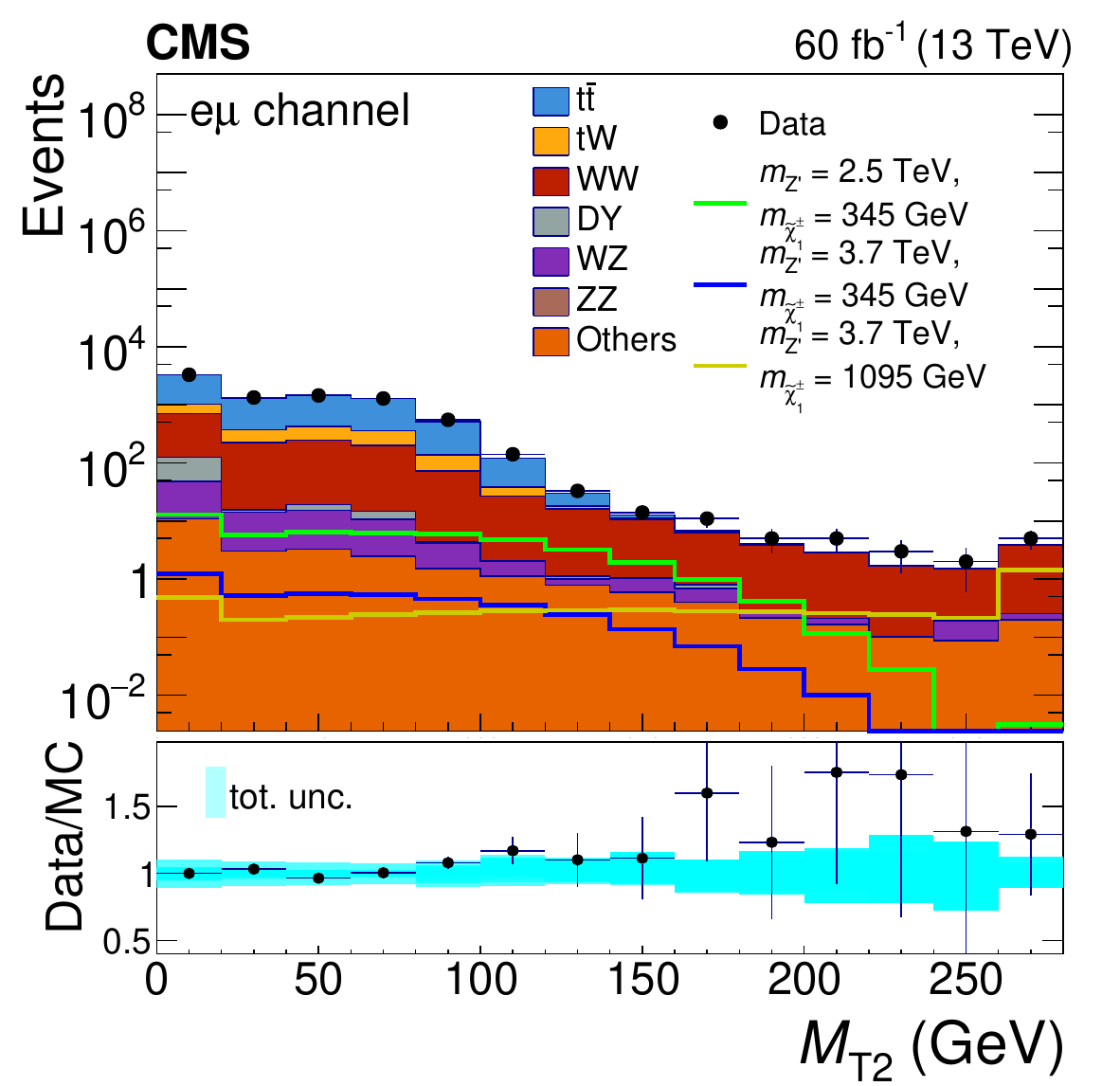}
\caption{Distributions measured from the \epmmmp sample collected in 2018, 
for $\pt(\Pell_1)$ (upper left), $m_{\Pell\Pell}$ (upper right), $\mT$ (lower left), and $M_\mathrm{T2}$ (lower right). 
Several benchmark signal distributions are overlaid (colored lines), illustrating the separation power between signal and background. 
The panel under each plot shows the data-to-background ratio,
along with the corresponding total uncertainty band (in cyan).}   
\label{fig:input_features}
\end{figure}

While the set of the most important variables slightly depends on the search channel and assumed signal scenario,
the top-ranked variables typically include $\pt(\Pell)$, $m_{\Pell\Pell}$, \ptmiss, $\mT$, and $M_\mathrm{T2}$.
Figure~\ref{fig:input_features} shows the distributions measured for some variables.

The analysis uses the TensorFlow~\cite{tensorflow2015-whitepaper} framework, interfaced with the Keras~\cite{chollet2015keras} library. 
The gradient descent algorithm~\cite{ruder2016overview} employed in the PNN assumes that the input variables follow normal Gaussian distributions, 
while, in reality, they span a wide variety of numerical ranges.
Therefore, each variable $x$ is transformed into $(x-\bar{x}) / \sigma_x$, 
where $\bar{x}$ and $\sigma_x$ are the mean and standard deviation of its distribution.
Model structure parameters (e.g., the number of hidden layers) and training process parameters
are scanned in a grid search and by using Bayesian optimization methods~\cite{7352306}.

\begin{figure}[ht]
\centering
\includegraphics[width=0.5\textwidth]{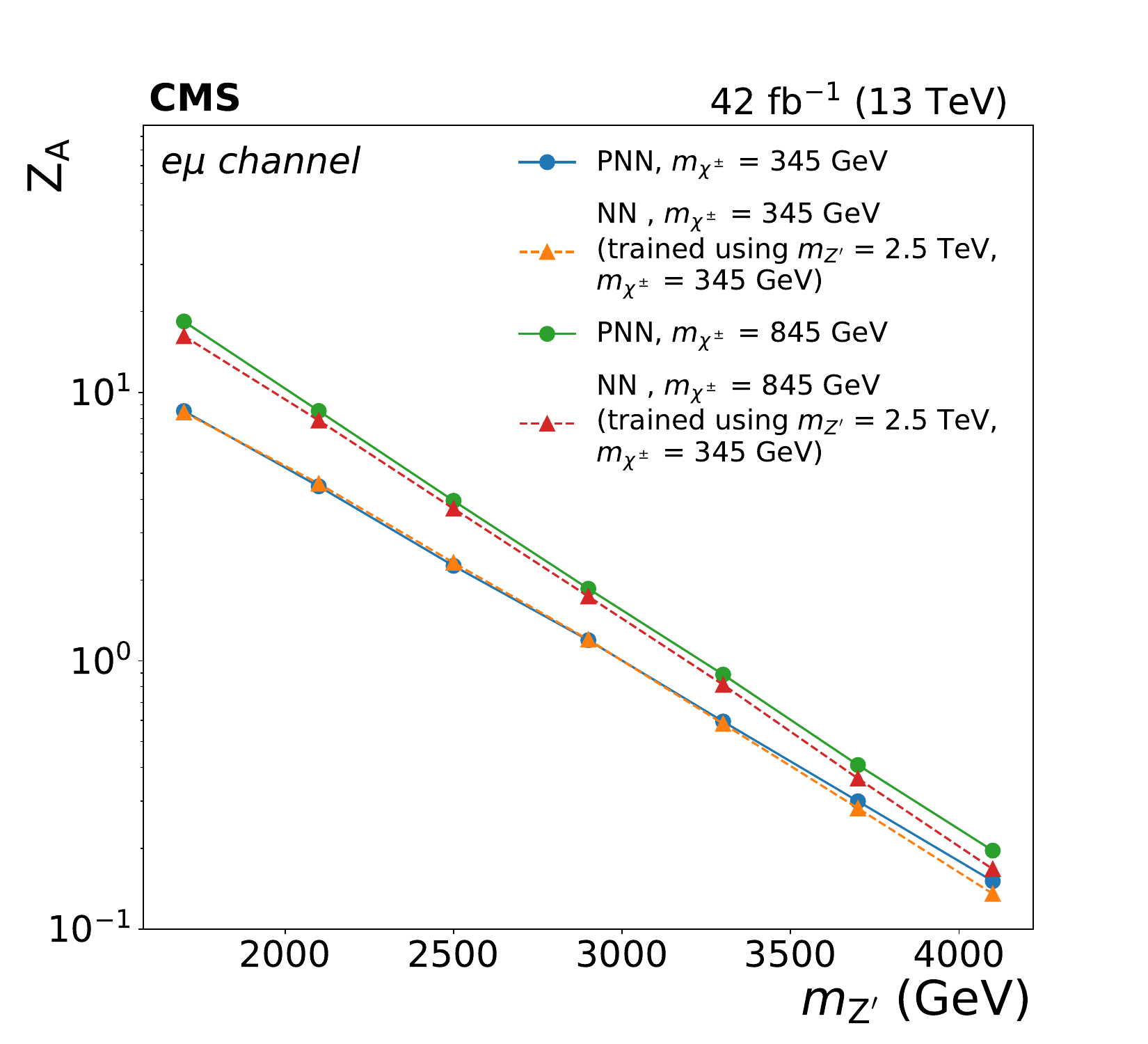}
\caption{Asimov significance, $Z_A$, vs.\ $m_\PZpr$ for the PNN model (circles) and for a NN model trained on a specific signal point (triangles) for the 2017 simulated event sample, in the \epmmmp channel. 
Two chargino masses have been considered, as indicated.}
\label{fig:pnn_vs_onemass}
\end{figure} 

Figure~\ref{fig:pnn_vs_onemass} shows the performance of the PNN approach with that of 
an NN model trained on a specific signal point ($m_\PZpr = 2.5$\TeV and $m_\PSGcpDo = 345$\GeV), 
using MC simulations of the $\Pe\PGm$ search channel with the 2017 integrated luminosity.
The comparison uses the Asimov significance ($Z_A$)~\cite{Cowan:2010js}, defined as

\begin{equation}
\begin{aligned}
Z_A^2 
&= -2 \ln \frac{\mathcal{L}(\mu=0 \mid D(s+b))}{\mathcal{L}(\mu=1 \mid D(s+b))}\\
&= -2 \ln \frac{\prod_i^\text{bins} \mathcal{P}(s_i+b_i \mid b_i)}{\prod_i^\text{bins} \mathcal{P}(s_i+b_i \mid s_i+b_i)}\\
&= \sum_i^\text{bins} 2 \Big[ (s_i+b_i) \ln \frac{s_i+b_i}{b_i} - s_i \Big] .
\end{aligned}
\end{equation}
where $\mu$ represents the signal strength (the ratio of the signal yield to the predicted signal yield, where $\mu=1$ corresponds to the full prediction), 
$D(s+b)$ denotes pseudo-data assuming the full signal and background yields, 
$\mathcal{P}$ represents the Poisson distribution, 
and $s_i$ ($b_i$) is the number of signal (background) counts in the $i^\text{th}$ bin of the PNN scores. 
The two significances coincide at the signal point used in the NN training ($m_\PZpr = 2.5$\TeV and $m_\PSGcpDo = 345$\GeV), as expected, 
but the NN significance degrades as the masses deviate from that point.
For example, when $m_\PZpr = 4.1$\TeV, the PNN significance is higher by 11 (17)\% for a chargino mass of 345 (845)\GeV.

\section{Background evaluation}

The dominant irreducible SM backgrounds in the signal region (SR) are the \ttbar, $\PW\PW$, and DY processes, whose shapes in the signal region are evaluated using the simulated events for each process.
These backgrounds are normalized using factors obtained by fitting the PNN score distributions in the SR and,
simultaneously, in three control regions (CRs) defined so that they are dominated by each of the three SM processes, 
with negligible signal contamination. 
The three search channels are analyzed separately and then combined in a fit.
The \ttbar CR is considered in all three channels, 
the $\PW\PW$ CR is only considered in the \epmmmp channel as its yield is negligible in the other channels,
and the DY CR is only considered in the \epem and \mpmm channels.

\begin{table}[ht]
\centering
\topcaption{Definitions of the signal and control regions used in the analysis.}
\label{tab:regiondef}
\begin{tabular}{ccccc} 
Region & $\ptmiss$ (\GeVns) & $m_{\Pell\Pell}$ (\GeVns) &  \PQb$\text{ jets}$ & Channels \\ 
\hline
SR    & $>$100 & $>$100 & 0 & \epem, \mpmm, \epmmmp \\ 
\ttbar CR & ${>}100$ & ${>}100$ & ${\ge}1$ & \epem, \mpmm, \epmmmp \\ 
$\PW\PW$ CR& ${>}100$ & 40--100  & 0 & \epmmmp \\ 
DY CR & 30--100 & ${>}100$  & 0 & \epem, \mpmm \\ 
\end{tabular}
\end{table}

Table~\ref{tab:regiondef} lists the selection criteria that define the SRs and CRs. 
Events with at least one \PQb-tagged jet are used to define the \ttbar CR, 
while those with $40 < m_{\Pell\Pell} < 100\GeV$ define the $\PW\PW$ CR. 
The DY CR is defined by requiring \ptmiss to be between 30 and 100\GeV. 
The same PNN model is deployed in all regions.

\begin{figure}[ht]
\centering
\includegraphics[width=0.4\textwidth]{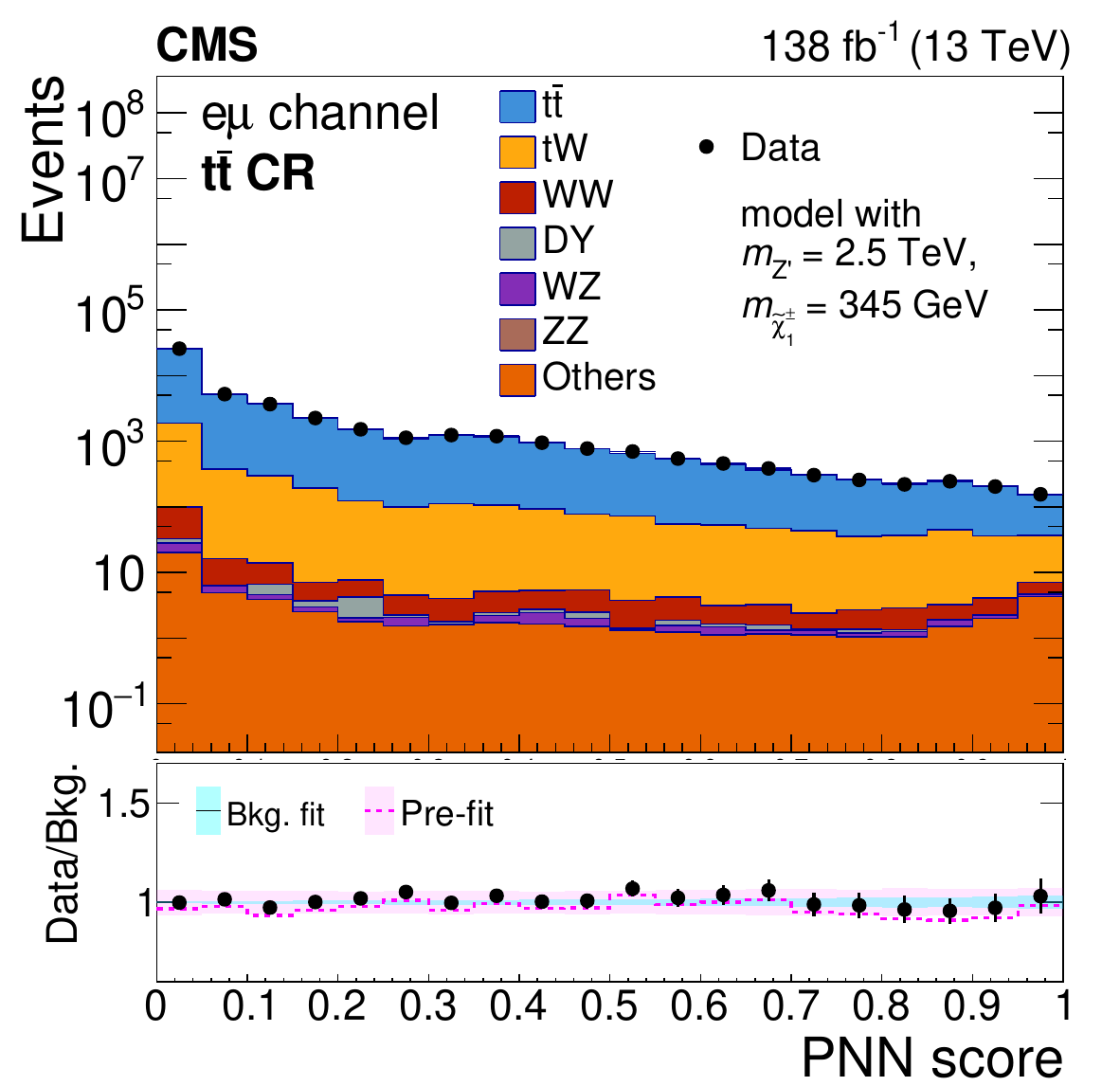}
\includegraphics[width=0.4\textwidth]{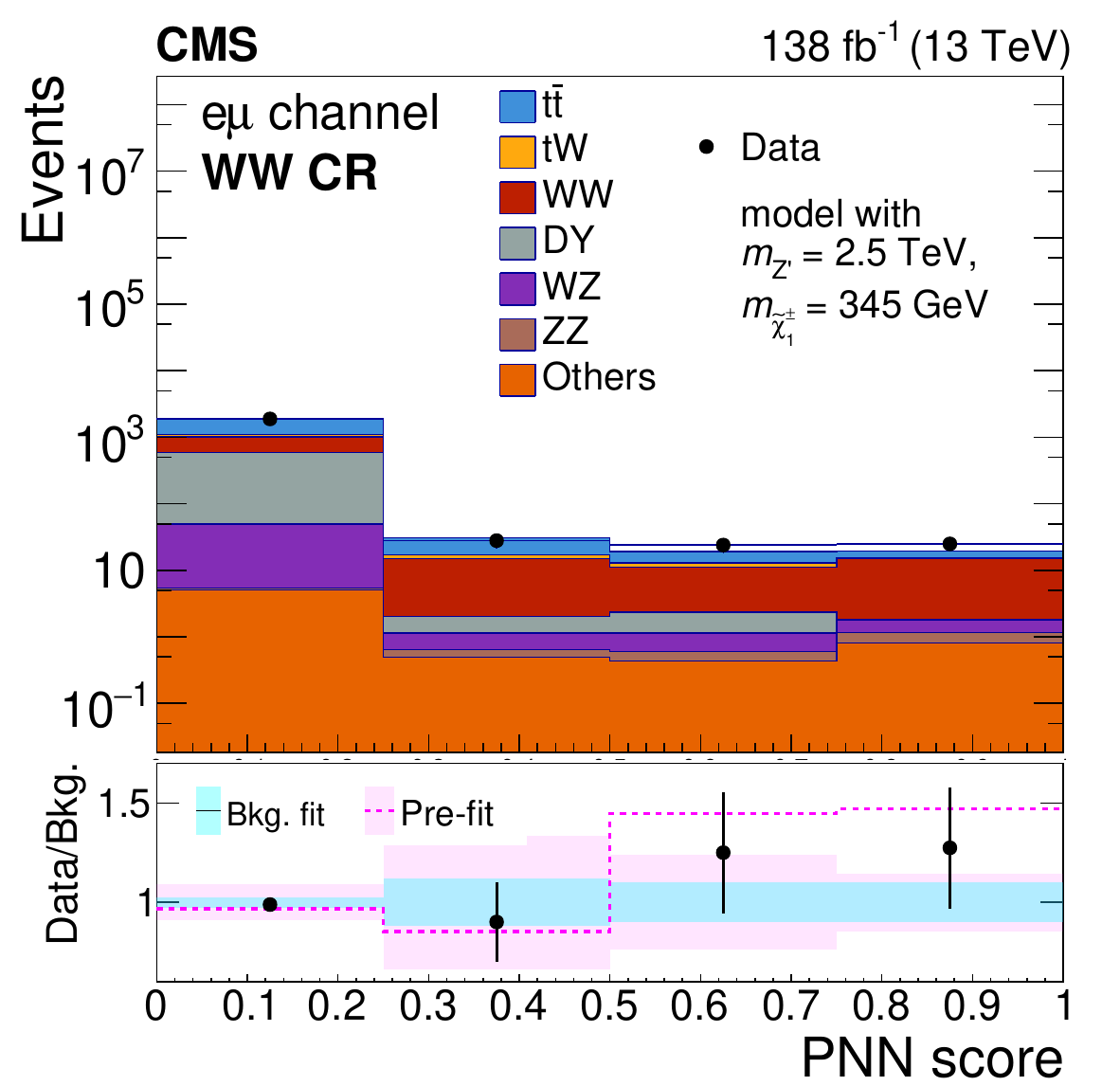}
\includegraphics[width=0.4\textwidth]{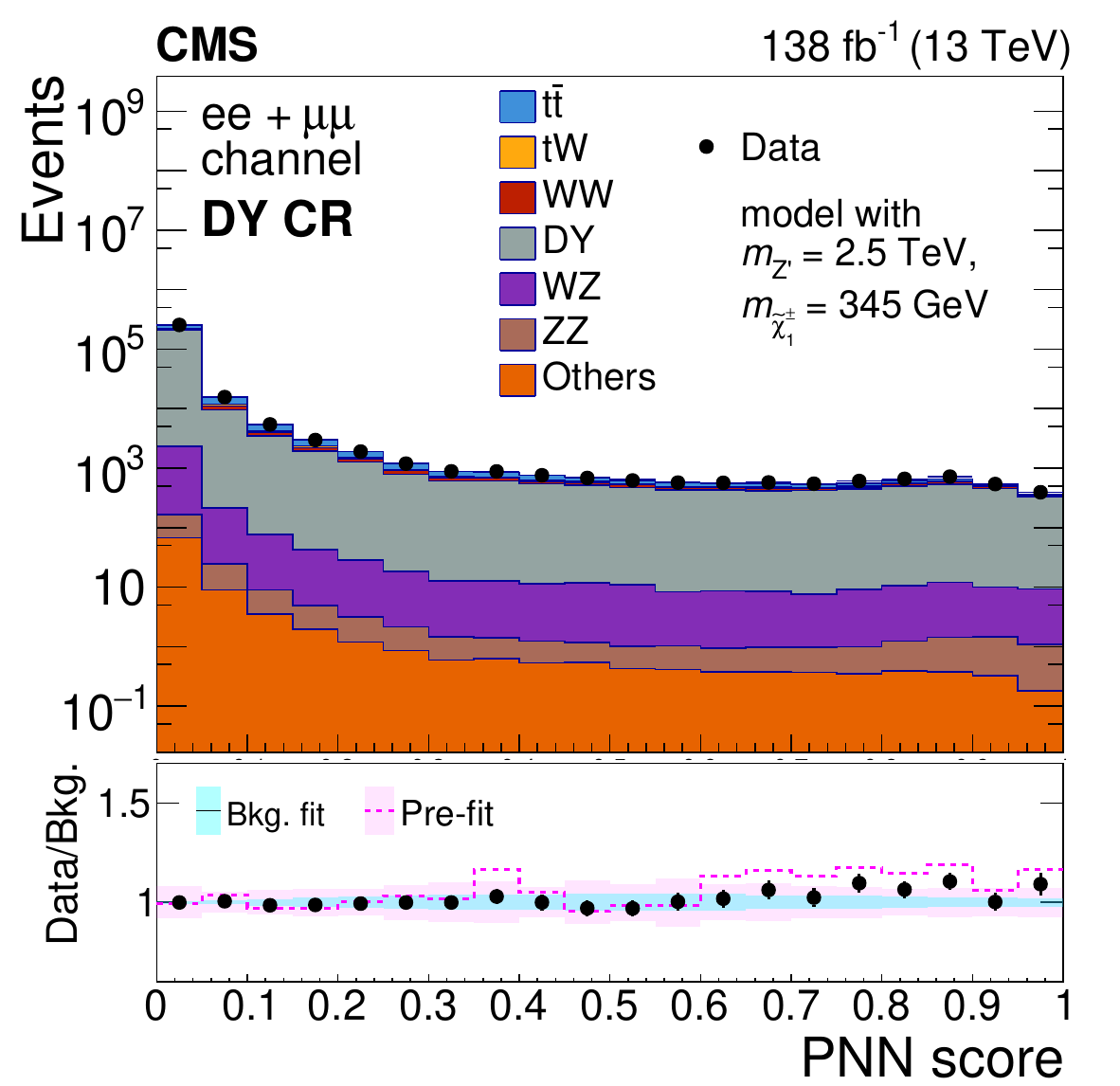}  
\caption{Measured and simulated SM PNN score distributions for a model with $m_{\PZpr} = 2.5$\TeV and $m_{\PSGcpmDo} = 345$\GeV, 
for the \ttbar (upper left) and $\PW\PW$ (upper right) CRs in the \epmmmp channel and 
for the DY CR in the $\epem+\mpmm$ channel (lower).
The panel under each plot shows the data-to-background ratios 
using either a background-only fit (black circles and cyan band) 
or a pre-fit (magenta dashed line and band). The uncertainties are displayed around the unity line.}
\label{fig:dnn_cr_channel}
\end{figure}

Figure~\ref{fig:dnn_cr_channel} shows the PNN score distributions of the measured 
(merging the three data-taking periods) and simulated SM background events,
for the PNN model with $m_{\PZpr} = 2.5$\TeV and $m_{\PSGcpmDo} = 345$\GeV,
for three combinations of CRs and search channels:
$\ttbar$ and $\PW\PW$ CRs for \epmmmp and DY CR for the merged \epem and \mpmm channels. 
The SM distributions in the colored stack show the fitted yields (``Bkg.\ fit").
The data-to-background ratios are shown using both the raw (``Pre-fit") and fitted SM yields.
The measured distributions are in agreement with the fitted SM backgrounds.

The remaining, subdominant, background processes have a negligible impact in the analysis sensitivity
and, therefore, are evaluated from the simulated samples.

\section{Systematic uncertainties}

Several sources of theoretical and experimental systematic uncertainties listed below
affect the shape or normalization of the PNN signal and background distributions.
These uncertainties are taken into account through nuisance parameters in the final profile likelihood fit. 
Effects from a single source of uncertainty are treated as fully correlated across the PNN score bins, 
whereas independent sources of uncertainty are treated as uncorrelated. 

The uncertainties associated with the renormalization ($\mu_{\text{R}}$) and factorization ($\mu_{\text{F}}$) scales 
are evaluated by independently varying $\mu_{\text{R}}$ and $\mu_{\text{F}}$ by a factor of 2 and 0.5, excluding the two extreme combinations.
The uncertainty associated with the PDFs and \alpS value is evaluated by reweighting the simulated events using the variations provided by the NNPDF3.1~\cite{Ball:2017nwa} and PDF4LHC~\cite{Butterworth:2015oua} sets, respectively, 
for the SM backgrounds and signal.
The uncertainty from the matching between the matrix element generator and the parton shower in the \POWHEG and \PYTHIA simulation 
is evaluated with simulation studies, varying the parameter ($h_\text{damp}$) that regulates the first high-\pt emission. 
Uncertainties in the simulation of the underlying event are estimated by varying the tune settings~\cite{CMS:2019csb}.
To improve the agreement between the measured and simulated top quark \pt spectra in \ttbar events, 
weights are applied to the NLO events to match the \pt distribution predicted at next-to-next-to-leading order (NNLO)~\cite{Czakon:2017wor}. 
The resulting variation in the prediction after applying the reweighting is taken as the corresponding uncertainty.
The backgrounds normalized using factors obtained from the fit (\ttbar, $\PW\PW$, and DY) are assigned a normalization uncertainty corresponding to the fit uncertainty. The backgrounds directly estimated from the simulated samples (all but \ttbar, $\PW\PW$, and DY) 
have a normalization uncertainty equal to twice their theoretical uncertainty.

The integrated luminosities of the 2016, 2017, and 2018 data taking periods have uncertainties 
of 1.2\%, 2.3\%, and 2.5\%, respectively~\cite{CMS-LUM-17-003,CMS-PAS-LUM-17-004,CMS-PAS-LUM-18-002}. 
These uncertainties are uniformly applied to the simulated events.
The number of simulated pileup interactions is estimated assuming a total inelastic $\Pp\Pp$ cross section of 69.2\unit{mb}. 
A corresponding uncertainty is evaluated by varying the total inelastic cross section by $\pm 4.6$\%~\cite{CMS:2020ebo}.
During the 2016 and 2017 data-taking periods, 
a gradual shift in the timing of the inputs of the ECAL L1 trigger in the forward endcap region ($\abs{\eta} > 2.4$) led to an inefficiency.
Also the muon detectors showed a similar trigger inefficiency effect, especially in 2016.
Corrections for this effect are applied to simulated events; the corresponding uncertainties are around 1\%.
Scale factors are applied, as functions of lepton \pt and $\eta$, 
to account for differences between the measured and simulated samples regarding the lepton reconstruction, identification, and isolation efficiencies. 
The uncertainties associated with the calculation of these factors are propagated as systematic uncertainties.
Uncertainties in the evaluation of the jet energy scale are taken into account by varying the jet momenta in simulated events, 
separately for the overall energy scale, differences in flavor response, residual differences between the energy scale measurements, etc. 
Uncertainties in the jet energy resolution are evaluated by varying the difference between the reconstructed and particle-level jet energies, 
or by smearing the measured jet energy if no matching particle-level jet could be found~\cite{Khachatryan:2016kdb}. 
These jet energy scale and resolution variations are also propagated to \ptvecmiss.
Differences between the measured and simulated \PQb jet tagging and light-quark or gluon jet mistagging scale factors 
are assigned as systematic uncertainties, as functions of \pt and $\eta$.
Statistical uncertainties resulting from the finite size of the simulated samples are addressed using the Barlow--Beeston method~\cite{Barlow:1993dm}:
in bins with more than 100 effective events a single Gaussian-constrained nuisance parameter is introduced for the sum of process yields,
while for the other bins Poisson-constrained parameters are introduced for each process.

Among the systematic sources, the dominant contributions arise from the normalization uncertainty for backgrounds normalized using fit factors (around 12\% of the fit uncertainty) and from the \PQb jet tagging scale factor uncertainty (around 10\% of the prior uncertainty). The other sources have a relatively mild impact on the results.

\section{Results and interpretation}

The search for the \PZpr signal is performed by fitting the measured PNN score distributions, 
in the SR for each of the three search channels.
These distributions depend on the assumed \PZpr and \PSGcpmDo masses.
The CRs are combined and fitted simultaneously with the SRs, 
to determine the normalization factors for the dominant background processes. 

\begin{figure}[ht]
\centering
\includegraphics[width=0.43\textwidth]{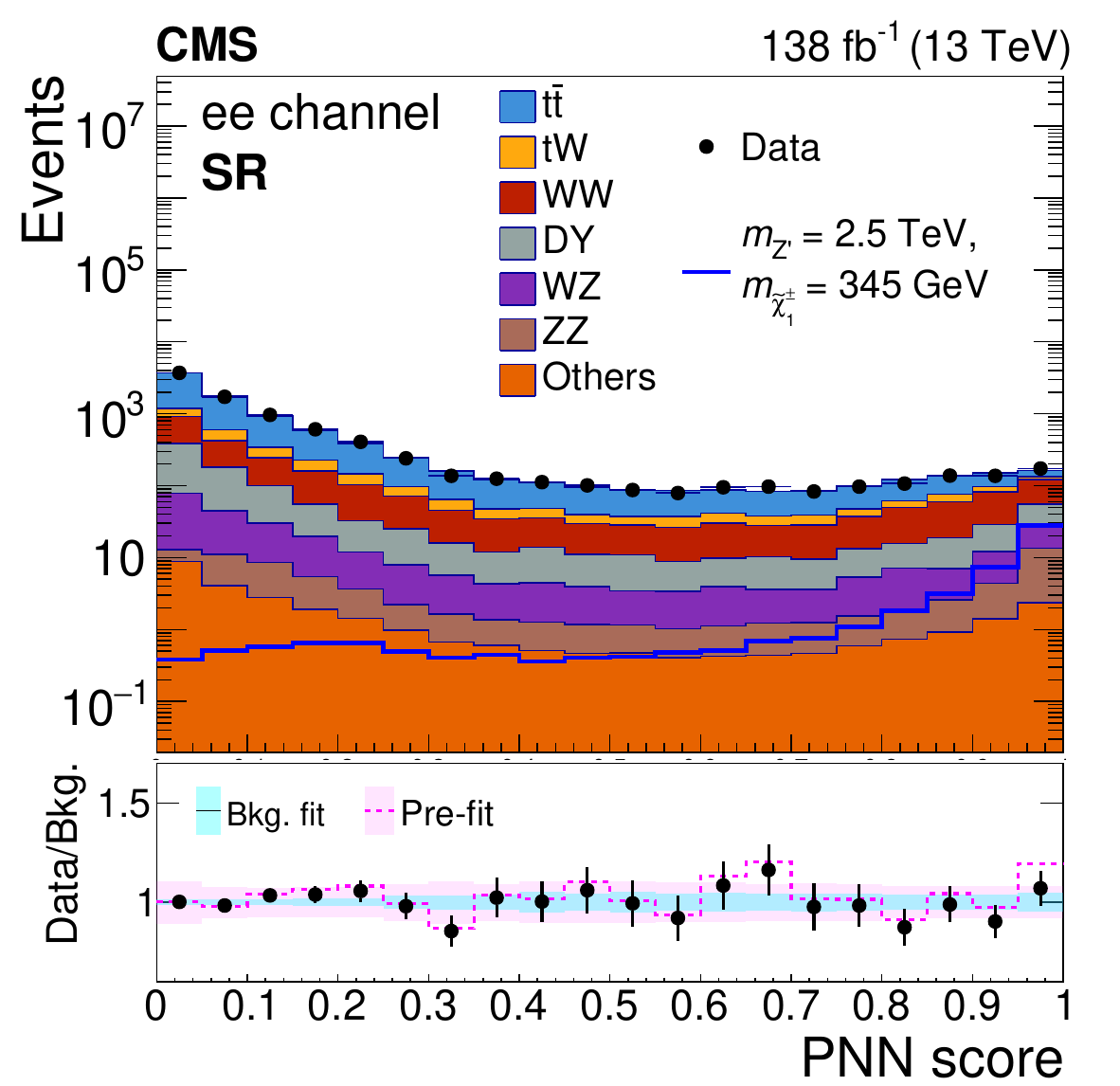} 
\includegraphics[width=0.43\textwidth]{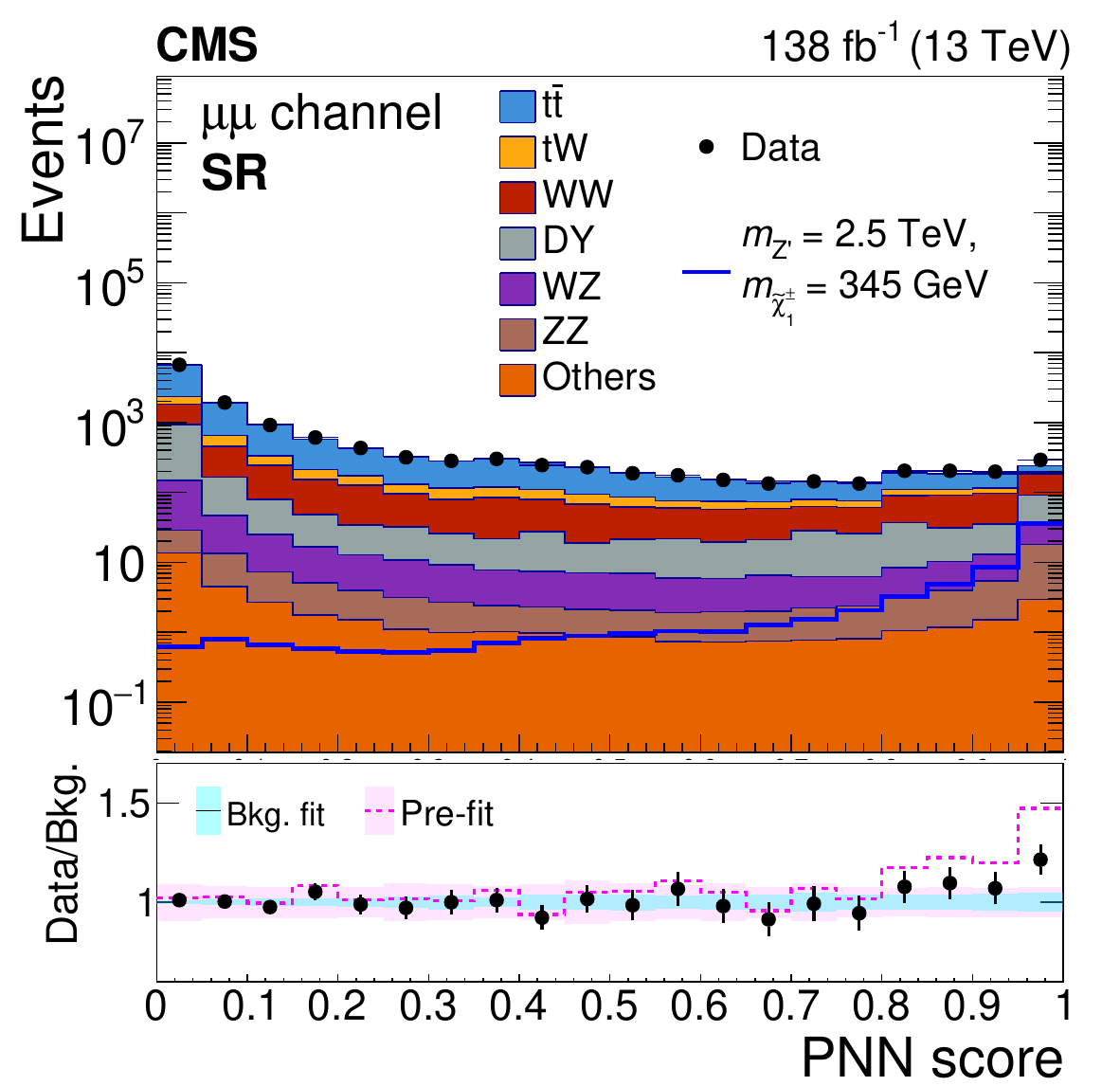} 
\includegraphics[width=0.43\textwidth]{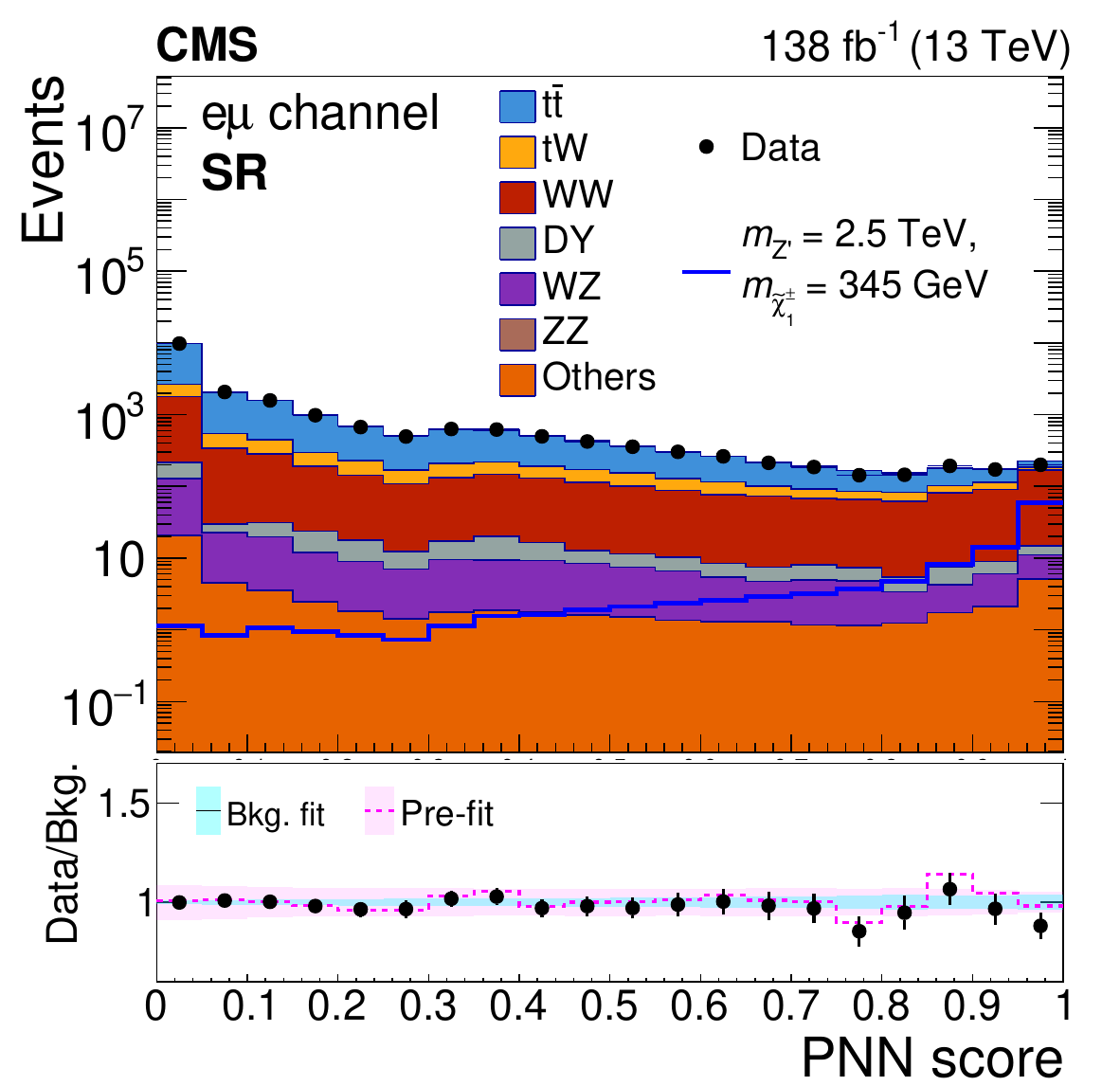} 
\caption{Measured and estimated background PNN score distributions, in the SRs of the three search channels.
Signal distributions for a model with $m_{\PZpr} = 2.5$\TeV and $m_{\PSGcpmDo} = 345$\GeV 
are superimposed on the plots.
The panel under each plot shows the data-to-background ratios 
using either a background-only fit (black circles and cyan band) 
or a pre-fit (magenta dashed line and band). The uncertainties are displayed around the unity line.}
\label{fig:dnn_sr_channel}
\end{figure}

Figure~\ref{fig:dnn_sr_channel} presents the measured and estimated background PNN score distributions in the SR of each of the three search channels,
for a model with $m_{\PZpr} = 2.5$\TeV and $m_{\PSGcpmDo} = 345$\GeV. 
In each panel, the stack of SM background contributions corresponds to the background-only fit. 
The panels under each plot show the data-to-background ratios in both the ``pre-fit" and ``background-only fit" options, 
also displaying the total uncertainties, computed as the sum in quadrature of the statistical and systematic terms. 
The normalization factors of the dominant background processes are primarily determined from the CRs 
and the background-rich bins in the SRs. 
For the signal mass point hypothesis shown in Fig.~\ref{fig:dnn_sr_channel},
the fitted normalization factors are: $1.01 \pm 0.03$ for \ttbar, $1.12 \pm 0.13$ for $\PW\PW$, and $1.14 \pm 0.04$ for DY.
The uncertainties (from the fit) are smaller for the processes with large yields in the CRs.

In the \epem channel, the fit shows agreement between the measured PNN distribution and the SM expectation.
In the \mpmm channel, a mild excess is observed in the signal-like bin (PNN score close to 1), 
with a local significance of 2 to 3 standard deviations, 
depending on the assumed signal mass hypothesis. 
The \epmmmp channel, instead, shows a mild deficit in the signal-like bin.

\begin{figure}[ht]
\centering
\includegraphics[width=0.65\textwidth]{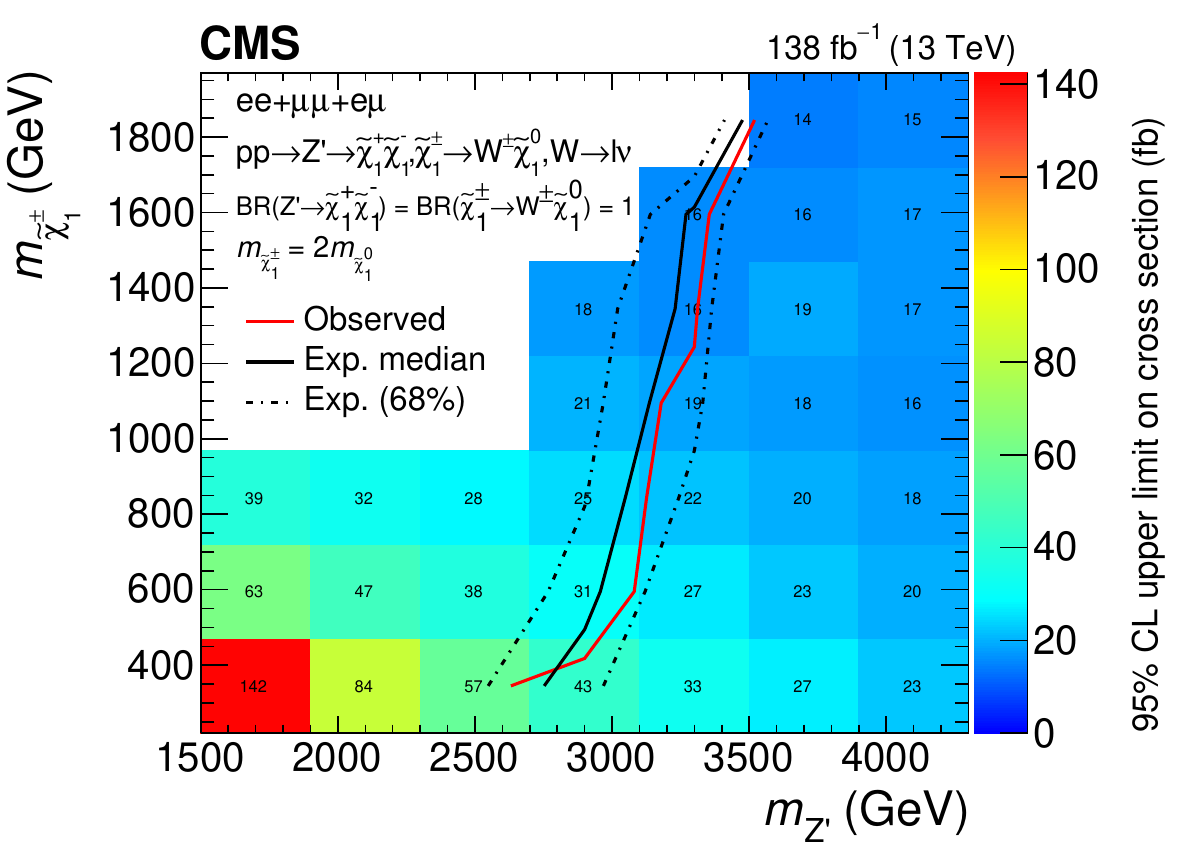}  
\caption{Upper limits, at 95\% \CL, on the $\Pp\Pp \to \PZpr$ cross section, 
in the \PZpr boson mass vs.\ \PSGcpmDo mass plane, 
combining the \epem, \mpmm, and \epmmmp channels,
as determined from the measured data (red line) 
and as expected from the simulation studies (black lines: median exclusion in solid and 68\% quantiles in dashed).
The region to the left of the curves is excluded.
The color grid shows median expected upper limits.}
\label{fig:limit_2d}
\end{figure}

\begin{figure}[ht]
\centering
\includegraphics[width=0.65\textwidth]{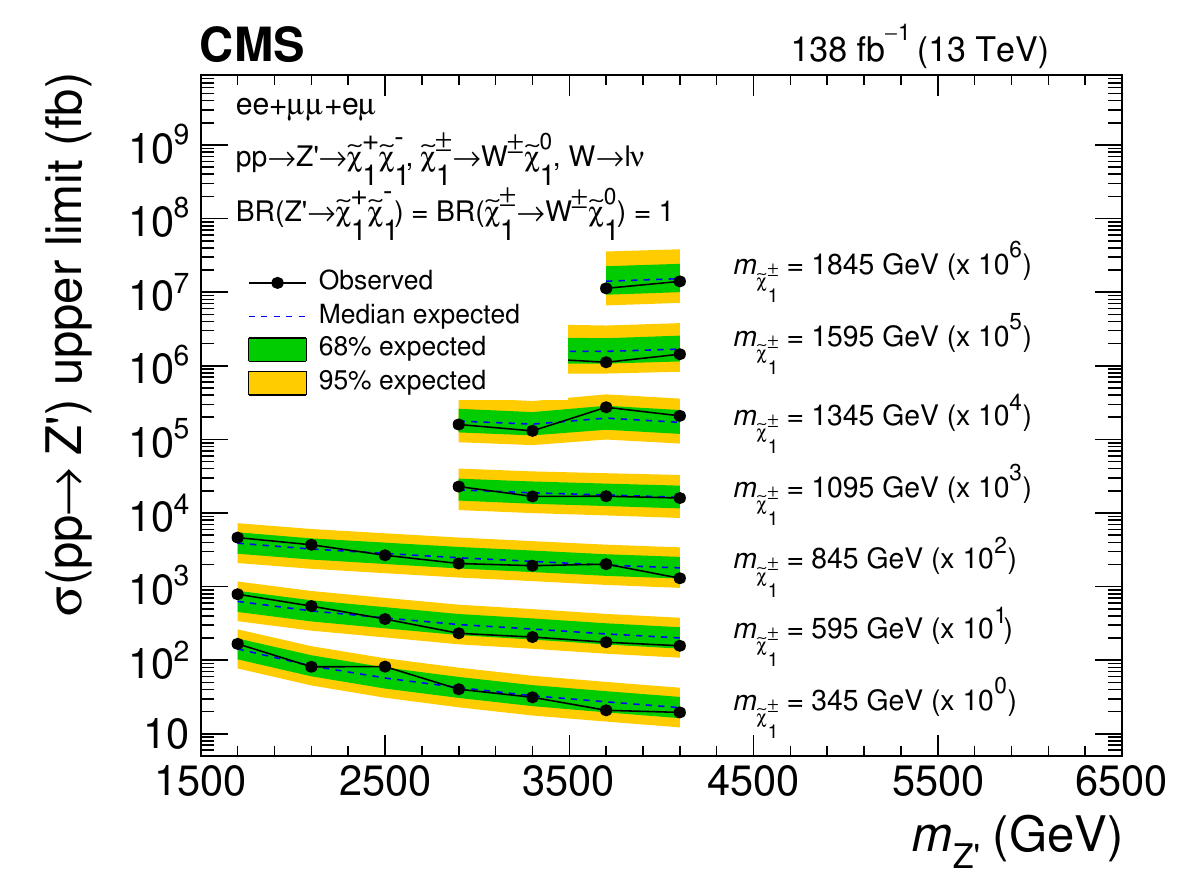}
\caption{Upper limits, at 95\% \CL, on the $\Pp\Pp \to \PZpr$ cross section,
vs.\ \PZpr boson mass for several \PSGcpmDo mass values,
combining the \epem, \mpmm, and \epmmmp channels,
as determined from the measured data (black circles) 
and as expected from the simulation studies (dashed lines with green and yellow uncertainty bands).
The limits for different $m_\PSGcpmDo$ values are shifted vertically, for visibility purposes.} 
\label{fig:limit_1d}
\end{figure}

A statistical interpretation has been conducted by testing the SM hypothesis against the \PZpr boson hypothesis. 
Upper limits on the \PZpr boson production cross section are calculated at 95\% confidence level (\CL) 
using a modified frequentist approach and the \CLs criterion, 
implemented through an asymptotic approximation~\cite{Cowan:2010js,Junk:1999kv,Read:2002hq,CMS-NOTE-2011-005}. 
The limits are computed using the \textsc{Combine} statistical analysis tool~\cite{CMS:2024onh}, 
which is based on the \textsc{RooFit}~\cite{Verkerke:2003ir} and \textsc{RooStats}~\cite{Moneta:2010pm} frameworks.
Figure~\ref{fig:limit_2d} shows the observed and expected $\Pp\Pp \to \PZpr$ cross section upper limits 
in the $m_\PZpr$ vs.\ $m_\PSGcpmDo$ plane, 
while Fig.~\ref{fig:limit_1d} shows the corresponding limits as a function of $m_\PZpr$, 
for several $m_\PSGcpmDo$ values.

The cross section limits are computed for an integrated luminosity of 138\fbinv and 
assuming that the $\PZpr \to \PSGcpDo\PSGcmDo \to \PWp \PSGczDo\PWm \PSGczDo$ branching fraction is 100\%.
The \PZpr boson width is considered negligible and the neutralino mass is set to be half of the chargino mass.
The obtained limits are mostly driven by the \epmmmp channel, 
which has a better sensitivity because its signal (and background) event yields are higher than those of the same-flavor channels. 
For a fixed \PZpr boson mass, the sensitivity improves for higher \PSGcpmDo masses, given the improved signal selection efficiency. 
Although the sensitivity is quoted for a narrow-width \PZpr resonance, 
it is expected to be similar for wider resonances with a width-to-mass ratio up to 20\%, as the kinematic distributions are found to be comparable to those in the narrow-width case.

For the scenario with $m_\PSGcpmDo \approx 1.8$\TeV, 
the measurement excludes the existence of \PZpr bosons with mass up to 3.5\TeV. 
Assuming that the \PZpr boson has a mass of 2.9\TeV and only decays to a chargino pair,
the observations exclude charginos in the 400--1400\GeV mass range.

Tabulated results are provided in the HEPData record for this analysis~\cite{hepdata}.

\section{Summary}
\label{sec:summary}

A search has been conducted for a leptophobic \PZpr boson decaying into two charginos, 
which subsequently decay into \PW bosons and neutralinos.  
A data sample of proton-proton collisions at a center-of-mass energy of 13\TeV, 
corresponding to an integrated luminosity of 138\fbinv, was analyzed.  
This is the first search for this process using LHC data. 
An analysis was designed in the dilepton plus missing transverse momentum final state, 
in the \epem, \mpmm, and \epmmmp channels. 
A parametrized neural network was employed to enhance the signal sensitivity. 
The analysis was interpreted using simplified model spectra 
featuring the production and decay process of the leptophobic \PZpr boson 
derived from the $U(1)^\prime$ extension of the minimal supersymmetric standard model.  The measurements were found to be consistent with the standard model expectations. 
Upper limits on the \PZpr boson production cross section were presented in the \PZpr boson mass vs.\ chargino mass plane, 
with $m_\PSGcpmDo = 2 \, m_\PSGczDo$.  
The analysis excludes \PZpr boson masses up to about 3.5\TeV 
for the specific case of \PZpr bosons decaying exclusively to charginos, with the charginos decaying to \PW bosons and neutralinos. Under the assumption of 2.9\TeV \PZpr boson decaying exclusively to charginos, observations rule out chargino masses in the 400--1400\GeV range. 

\begin{acknowledgments}
We congratulate our colleagues in the CERN accelerator departments for the excellent performance of the LHC and thank the technical and administrative staffs at CERN and at other CMS institutes for their contributions to the success of the CMS effort. In addition, we gratefully acknowledge the computing centers and personnel of the Worldwide LHC Computing Grid and other centers for delivering so effectively the computing infrastructure essential to our analyses. Finally, we acknowledge the enduring support for the construction and operation of the LHC, the CMS detector, and the supporting computing infrastructure provided by the following funding agencies: SC (Armenia), BMBWF and FWF (Austria); FNRS and FWO (Belgium); CNPq, CAPES, FAPERJ, FAPERGS, and FAPESP (Brazil); MES and BNSF (Bulgaria); CERN; CAS, MoST, and NSFC (China); MINCIENCIAS (Colombia); MSES and CSF (Croatia); RIF (Cyprus); SENESCYT (Ecuador); ERC PRG and PSG, TARISTU24-TK10 and MoER TK202 (Estonia); Academy of Finland, MEC, and HIP (Finland); CEA and CNRS/IN2P3 (France); SRNSF (Georgia); BMFTR, DFG, and HGF (Germany); GSRI (Greece); NKFIH (Hungary); DAE and DST (India); IPM (Iran); SFI (Ireland); INFN (Italy); MSIT and NRF (Republic of Korea); MES (Latvia); LMTLT (Lithuania); MOE and UM (Malaysia); BUAP, CINVESTAV, CONACYT, LNS, SEP, and UASLP-FAI (Mexico); MOS (Montenegro); MBIE (New Zealand); PAEC (Pakistan); MES, NSC, and NAWA (Poland); FCT (Portugal); MESTD (Serbia); MICIU/AEI and PCTI (Spain); MOSTR (Sri Lanka); Swiss Funding Agencies (Switzerland); MST (Taipei); MHESI (Thailand); TUBITAK and TENMAK (T\"{u}rkiye); NASU (Ukraine); STFC (United Kingdom); DOE and NSF (USA).

\hyphenation{Rachada-pisek} Individuals have received support from the Marie-Curie program and the European Research Council and Horizon 2020 Grant, contract Nos.\ 675440, 724704, 752730, 758316, 765710, 824093, 101115353, 101002207, 101001205, and COST Action CA16108 (European Union); the Leventis Foundation; the Alfred P.\ Sloan Foundation; the Alexander von Humboldt Foundation; the Science Committee, project no. 22rl-037 (Armenia); the Fonds pour la Formation \`a la Recherche dans l'Industrie et dans l'Agriculture (FRIA) and Fonds voor Wetenschappelijk Onderzoek contract No. 1228724N (Belgium); the Beijing Municipal Science \& Technology Commission, No. Z191100007219010, the Fundamental Research Funds for the Central Universities, the Ministry of Science and Technology of China under Grant No. 2023YFA1605804, the Natural Science Foundation of China under Grant No. 12535004, and USTC Research Funds of the Double First-Class Initiative No.\ YD2030002017 (China); the Ministry of Education, Youth and Sports (MEYS) of the Czech Republic; the Shota Rustaveli National Science Foundation, grant FR-22-985 (Georgia); the Deutsche Forschungsgemeinschaft (DFG), among others, under Germany's Excellence Strategy -- EXC 2121 ``Quantum Universe" -- 390833306, and under project number 400140256 - GRK2497; the Hellenic Foundation for Research and Innovation (HFRI), Project Number 2288 (Greece); the Hungarian Academy of Sciences, the New National Excellence Program - \'UNKP, the NKFIH research grants K 131991, K 133046, K 138136, K 143460, K 143477, K 146913, K 146914, K 147048, 2020-2.2.1-ED-2021-00181, TKP2021-NKTA-64, and 2025-1.1.5-NEMZ\_KI-2025-00004 (Hungary); the Council of Science and Industrial Research, India; ICSC -- National Research Center for High Performance Computing, Big Data and Quantum Computing, FAIR -- Future Artificial Intelligence Research, and CUP I53D23001070006 (Mission 4 Component 1), funded by the NextGenerationEU program, the Italian Ministry of University and Research (MUR) under Bando PRIN 2022 -- CUP I53C24002390006, PRIN PRIMULA 2022RBYK7T (Italy); the Latvian Council of Science; the Ministry of Education and Science, project no. 2022/WK/14, and the National Science Center, contracts Opus 2021/41/B/ST2/01369, 2021/43/B/ST2/01552, 2023/49/B/ST2/03273, and the NAWA contract BPN/PPO/2021/1/00011 (Poland); the Funda\c{c}\~ao para a Ci\^encia e a Tecnologia (Portugal); the National Priorities Research Program by Qatar National Research Fund; MICIU/AEI/10.13039/501100011033, ERDF/EU, ``European Union NextGenerationEU/PRTR", projects PID2022-142604OB-C21, PID2022-139519OB-C21, PID2023-147706NB-I00, PID2023-148896NB-I00, PID2023-146983NB-I00, PID2023-147115NB-I00, PID2023-148418NB-C41, PID2023-148418NB-C42, PID2023-148418NB-C43, PID2023-148418NB-C44, PID2024-158190NB-C22, RYC2021-033305-I, RYC2024-048719-I, CNS2023-144781, CNS2024-154769 and Plan de Ciencia, Tecnolog{\'i}a e Innovaci{\'o}n de Asturias, Spain; the Chulalongkorn Academic into Its 2nd Century Project Advancement Project, the National Science, Research and Innovation Fund program IND\_FF\_68\_369\_2300\_097, and the Program Management Unit for Human Resources \& Institutional Development, Research and Innovation, grant B39G680009 (Thailand); the Eric \& Wendy Schmidt Fund for Strategic Innovation through the CERN Next Generation Triggers project under grant agreement number SIF-2023-004; the Kavli Foundation; the Nvidia Corporation; the SuperMicro Corporation; the Welch Foundation, contract C-1845; and the Weston Havens Foundation (USA).  

\end{acknowledgments}\section*{Data availability} Release and preservation of data used by the CMS Collaboration as the basis for publications is guided by the  \href{https://doi.org/10.7483/OPENDATA.CMS.1BNU.8V1W}{CMS data preservation, re-use and open access policy}.
\bibliography{auto_generated}

\cleardoublepage \appendix\section{The CMS Collaboration \label{app:collab}}\begin{sloppypar}\hyphenpenalty=5000\widowpenalty=500\clubpenalty=5000\cmsinstitute{Yerevan Physics Institute, Yerevan, Armenia}
{\tolerance=6000
A.~Hayrapetyan, V.~Makarenko\cmsorcid{0000-0002-8406-8605}, A.~Tumasyan\cmsAuthorMark{1}\cmsorcid{0009-0000-0684-6742}
\par}
\cmsinstitute{Institut f\"{u}r Hochenergiephysik, Vienna, Austria}
{\tolerance=6000
W.~Adam\cmsorcid{0000-0001-9099-4341}, J.W.~Andrejkovic, L.~Benato\cmsorcid{0000-0001-5135-7489}, T.~Bergauer\cmsorcid{0000-0002-5786-0293}, M.~Dragicevic\cmsorcid{0000-0003-1967-6783}, C.~Giordano\cmsorcid{0000-0001-6317-2481}, P.S.~Hussain\cmsorcid{0000-0002-4825-5278}, M.~Jeitler\cmsAuthorMark{2}\cmsorcid{0000-0002-5141-9560}, N.~Krammer\cmsorcid{0000-0002-0548-0985}, A.~Li\cmsorcid{0000-0002-4547-116X}, D.~Liko\cmsorcid{0000-0002-3380-473X}, M.~Matthewman, I.~Mikulec\cmsorcid{0000-0003-0385-2746}, J.~Schieck\cmsAuthorMark{2}\cmsorcid{0000-0002-1058-8093}, R.~Sch\"{o}fbeck\cmsAuthorMark{2}\cmsorcid{0000-0002-2332-8784}, D.~Schwarz\cmsorcid{0000-0002-3821-7331}, M.~Shooshtari\cmsorcid{0009-0004-8882-4887}, M.~Sonawane\cmsorcid{0000-0003-0510-7010}, W.~Waltenberger\cmsorcid{0000-0002-6215-7228}, C.-E.~Wulz\cmsAuthorMark{2}\cmsorcid{0000-0001-9226-5812}
% ...
\par}
\vskip\cmsinstskip
\dag:~Deceased\\
$^{1}$Also at Yerevan State University, Yerevan, Armenia\\
$^{2}$Also at TU Wien, Vienna, Austria\\
% ...
\end{sloppypar}
%%% END EDITABLE REGION %%%
% skeleton_end
\end{document}